# Anomalous Hall effect in electrolytically reduced PdCoO$_2$ thin films

Yiting Liu,[a,b,†] Gaurab Rimal,[b,†] Pratyankara Narasimhan,[b] Seongshik Oh[b,*]

[a] State Key Laboratory of Precision Spectroscopy, East China Normal University, Shanghai 200062, China

[b] Department of Physics & Astronomy, Rutgers, The State University of New Jersey, Piscataway, New Jersey 08854, USA

[†] These authors contributed equally to this work.

[*] Corresponding author: ohsean@physics.rutgers.edu

## Abstract

PdCoO$_2$, being highly conductive and anisotropic, is a promising material for fundamental and technological applications. Recently, reduced PdCoO$_2$ thin films were shown to exhibit ferromagnetism after hydrogen annealing. Here, we demonstrate that when PdCoO$_2$ film is used as a cathode for dissociation of water, hydrogen generated around the film reduces the material and leads to the emergence of anomalous Hall effect (AHE). Moreover, we demonstrate that the sign of the AHE signal can also be changed with the electrolytic process. Electrolytically-modified PdCoO$_2$ films may open a door to applications in spintronics.

## 1. Introduction

$PdCoO_2$ is a promising oxide with a unique quasi-two-dimensional electronic structure. It has alternating Pd and $CoO_2$ layers, with each layer having a triangular lattice. Contributed mainly by Pd layers, $PdCoO_2$ has room-temperature conductivity comparable to those of the most conductive metals, such as copper, silver and gold [1]. Although $PdCoO_2$ and other delafossites were first synthesized in 1971 [2,3], $PdCoO_2$ was largely ignored for decades. However, following the synthesis of high quality single crystals in recent years [4–6], there has been renewed interest in this system for the study of properties such as the thermoelectric power [7–9], electronic structure [10–12] and high anisotropy in structure, conductivity and compression behavior [13–15]. First-principles calculations suggest that the high in-plane conductivity of $PdCoO_2$ originates mostly from the Pd band [10]. Hicks *et al.* also presented high-resolution de Haas-van Alphen oscillations and obtained a long mean free path of ~20 μm at 10 K [16], which is the longest among any oxide. Subsequently, hydrodynamic transport was observed in this material, which allowed the determination of electronic viscosity [17].

Delafossites were also studied as part of electrodes and catalysts in water electrolysis [18–20]. Hinogami *et al.* investigated catalytic activities of copper delafossites, such as $CuFeO_2$ and $CuCoO_2$ [21,22]. Podjaski *et al.* found that Co dissolution in the electrolyte and Pd-rich surface layer in $PdCoO_2$ promotes electrolysis efficiency [18]. Li *et al.* demonstrated that upon electrolysis of $PdCoO_2$ bulk crystals, Pd nanoclusters and Co oxides form and lead to high-efficiency hydrogen evolution [19].

$PdCoO_2$ is only weakly paramagnetic, but shows signatures of surface magnetism on Pd terminated surfaces [23]. Harada *et al.* found thickness-dependent anomalous Hall effect in $PdCoO_2$ thin films [24], and more recently, Rimal *et al.* demonstrated that reduction of $PdCoO_2$ thin films via hydrogen annealing can exhibit strong ferromagnetism with perpendicular magnetic anisotropy (PMA) [25], which was mainly as a result of oxygen loss and subsequent transformation of $PdCoO_2$ film into an atomically-intermixed Pd-Co alloy. Here, we show that the electronic properties of $PdCoO_2$ films can be substantially modified via electrolysis. By using different cathode voltages for water splitting (i.e. hydrogen generation) and studying the evolution of Hall effect, we find that $PdCoO_2$ films are reduced, exhibit ferromagnetism, and show different signs of anomalous Hall effect (AHE) depending on the applied voltage.

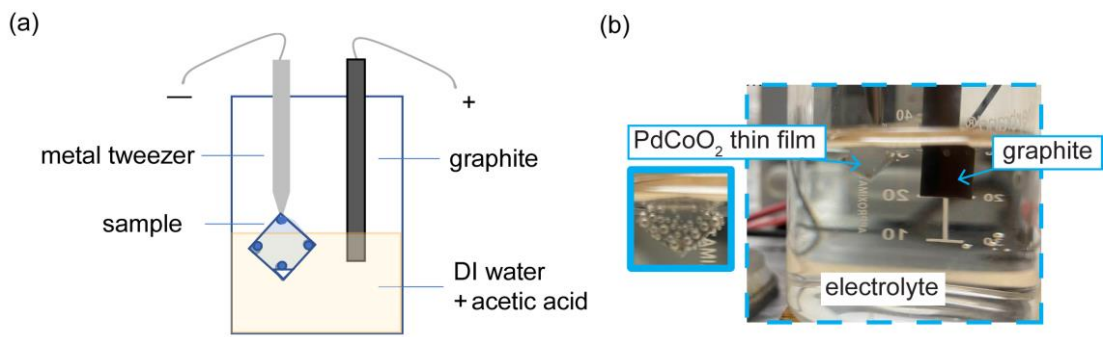

**Figure 1:** Experiment setup. (a) Schematic diagram of the setup; (b) Underwater part of the film. Also shown is the formation of hydrogen bubbles on the film surface when a voltage of -5.0 V is applied.

## 2. Experimental details

For this study, we utilized 9 nm thick $PdCoO_2$ thin films, grown on $10 \times 10$ mm$^2$, $Al_2O_3$ (0001) substrate using oxygen plasma-assisted molecular beam epitaxy with a base pressure of low $10^{-8}$ Pa [26]. After growth, the sample is cut into four pieces which are used for four separate experiments. Before each electrolysis experiment, indium dots are bonded to the sample corners as contact points for electrical measurements as shown in Figure 1(a,b). A metal tweezer holds the film just above the water level, while a graphite rod is used as the second electrode. Unless stated otherwise, $PdCoO_2$ is used as the negative electrode. We used 30 mL water with 2 drops of acetic acid as the electrolyte to facilitate hydrogen evolution. Electrolysis is powered by a variable voltage source. After each process, we measured the average two-probe resistance between indium contacts ($R_{2prb}$) on the corners of the film using a multimeter. Afterwards, Hall effect measurements were carried out using van der Pauw method. XPS was done on as-prepared samples using ThermoFisher K-alpha system using monochromated Al K-α beam (1486.7 eV). The XPS binding energy and resolution was calibrated using Ag $3d_{5/2}$ line on a clean Ag surface. Rutherford backscattering spectrometry (RBS) was performed at Rutgers ion scattering facility using 2 MeV $He^{2+}$ ions.

## 3. Results and discussion

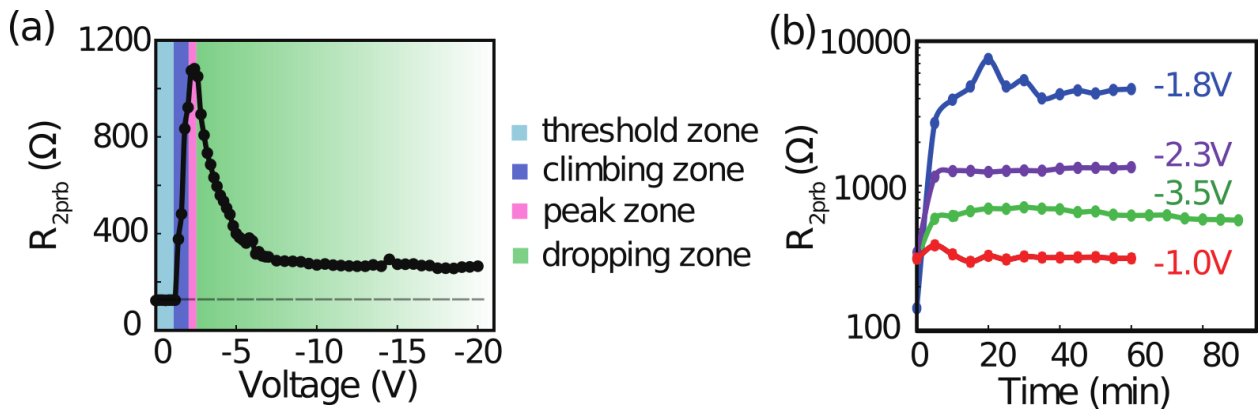

**Figure 2:** Average two-point resistance ($R_{2prb}$) measured after a sample is kept in the electrolyte (a) at the specified voltage for 5 minutes and (b) for the specified time at -1.0 V, -1.8 V, -2.3 V and -3.5 V. Four separate samples are used for (b). The negative voltages imply that the $PdCoO_2$ film is used as a cathode.

In Figure 2(a), we show how the applied voltage affects the film. Below -1.2 V, the film resistance remains unchanged, because no hydrogen is generated: note that this voltage is consistent with the well-known water splitting potential of -1.23 V. This region is termed the *threshold zone*. A steep increase occurs just above this threshold voltage and the resistance peaks at about -2.4 ± 0.2 V and slightly plateaus to about -2.6 V. The voltage range of -1.2 V to -2.2 V is thus termed the *climbing zone* and -2.2 V to -2.6 V is the *peak zone*. Past the peak zone, in the region termed *dropping zone*, the resistance starts to decrease gradually and reaches a stable value after -7.5 V. The resistance for the stable zone is higher than that of the original film. When we used $PdCoO_2$ film as an anode, the film remained unaffected up to the maximum applied voltage of 5.0 V.

In Figure 2(b), we present how the film resistance changes as a function of the process time for each of the four zones: *threshold* (-1.0 V), *climbing* (-1.8 V), *peak* (-2.3 V) and *dropping zone* (-3.5 V), respectively. While in the threshold zone (-1.0 V), the film resistance remains stable around the initial value, suggesting no changes in the film. This is because no hydrogen is evolved, so the film remains unaffected. For all the other zones the resistance rises fast initially and stabilizes after about 20 minutes. Interestingly, the stabilized resistance is highest at -1.8 V and decreases at higher voltages. This trend is similar to our earlier work on hydrogenation of $PdCoO_2$ films, in which the resistance initially rises with gradual oxygen removal and then drops after extended annealing [25].

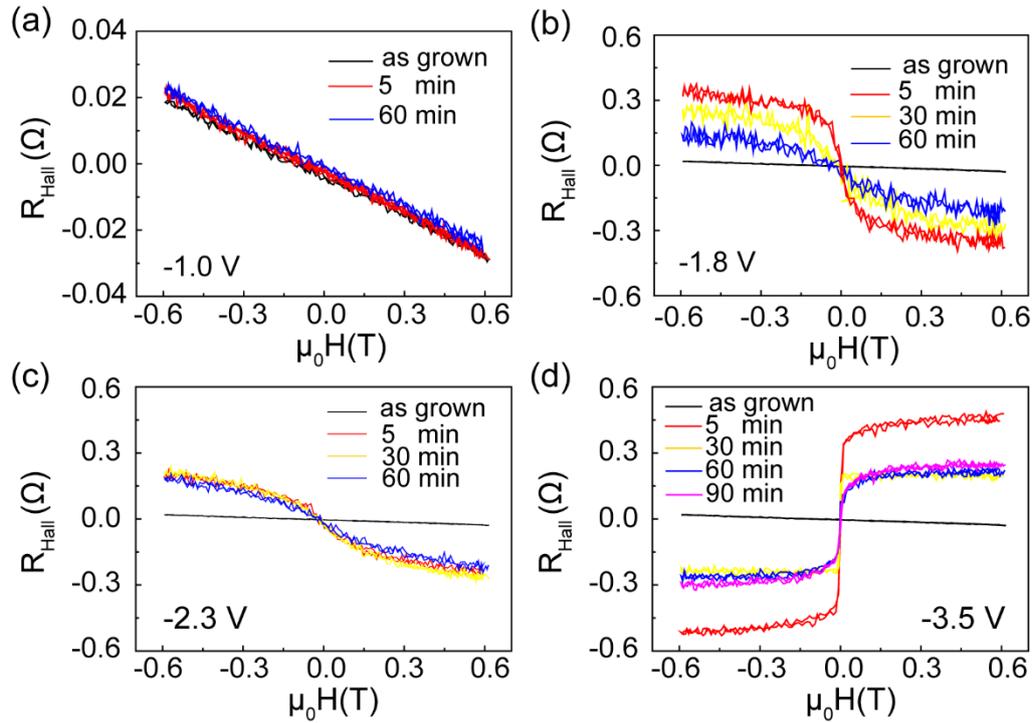

**Figure 3:** Hall effect data ($R_{Hall} \equiv R_{xy}$) of PdCoO$_2$ thin films measured at room temperature after electrolytic processing at (a) -1.0 V, (b) -1.8 V, (c) -2.3 V and (d) -3.5 V.

To determine the nature of the resulting films, we investigated their transport behavior. The room temperature Hall effect is shown in Figure 3. The as-grown PdCoO$_2$ film shows clear n-type linear behavior, and no change is observed for applied voltage of -1.0 V. Past this threshold voltage, AHE is present, as shown in Figure 3(b-d). It is notable that at the highest voltage (-3.5 V, Figure 3(d)), the AHE changes its sign to positive. This behavior suggests that the reaction kinetics play a large role in the overall properties of the film. The applied voltage controls redox activity [27], thereby the generated hydrogen, which subsequently changes the film properties. Similar observations were reported in ref. [25], where the reaction kinetics were shown to be important in driving the film properties.

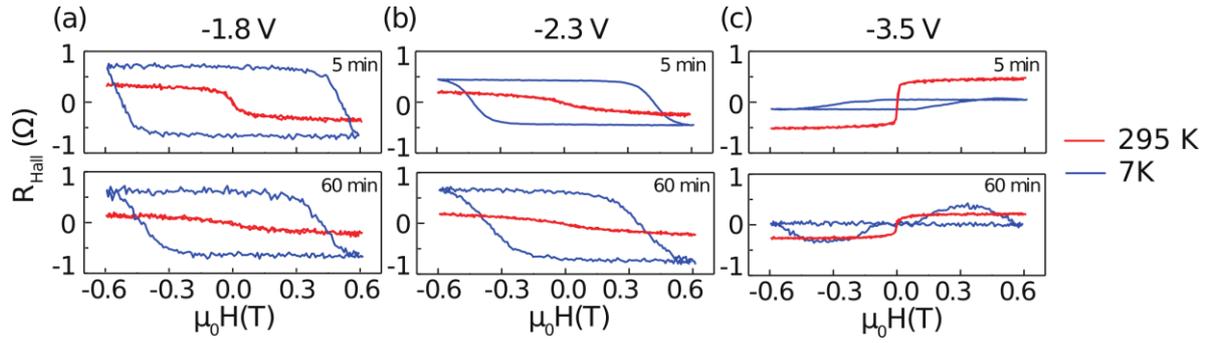

**Figure 4:** Hall resistance of PdCoO$_2$ films electrolytically processed for different duration at (a) -1.8 V, (b) -2.3 V, and (c) -3.5 V, measured at room temperature and 7 K.

In contrast to room temperature, Hall effect at 7 K, as shown in Figure 4, exhibits well-defined hysteresis loops with large coercive fields, suggesting the presence of stable ferromagnetic domains with PMA. Nonetheless, the magnitude of the coercive field gradually decreases as the voltage increases, with the sign switched at -3.5 V. A bowknot shape is observed after 60 min electrolysis at -3.5 V, which is likely due to the presence of multiple transport channels with different AHE signs. The AHE results are reproducible across multiple samples.

The emergence of AHE suggests that the main outcome of the electrolytic process is reduction of the PdCoO$_2$ films into a ferromagnetic Pd-Co alloy, which subsequently exhibits sign-tunable AHE with the applied voltage. The XPS data in Figure 5(a) confirm that the films are reduced after the process. It is notable that the electrolytic process, which is distinct from the hydrogen annealing [25], can induce similar modification of the PdCoO$_2$ films. There are some differences as well: most significantly, hysteresis is lacking at room temperature in these samples unlike the hydrogen-annealed samples. This suggests that despite some similarities, the electrolytic process, driven by an electro-chemical force, results in different structures (and compositions) of Pd-Co alloys than the hydrogen annealing does for the PdCoO$_2$ films. Notably, a recent electrolytic study of PdCoO$_2$ bulk crystals demonstrated that Co and O can leach into an acidic electrolyte leaving a Pd-rich capping layer [18]. In a similar manner, as shown in Figure 5(b), RBS comparing a pristine film and a film treated at -2.3 V for 60 mins shows significant loss of Co from the electrolytically treated film. This is in contrast with the hydrogen-annealed $PdCoO_2$ films, where Pd and Co contents remain intact through the process [25]. It is notable that even with the significant loss of Co, strong AHE signals are observed in the electrolytically-processed films. Considering that Pd is known to be on the verge of becoming ferromagnetic, even tiny amounts of Co can allow Pd to become ferromagnetic [28]. Moreover, Pd layer could also absorb hydrogen and form magnetic $PdH_x$ [29]. These studies suggest that ferromagnetism, thus AHE, can emerge via multiple routes during electrolysis in $PdCoO_2$ films.

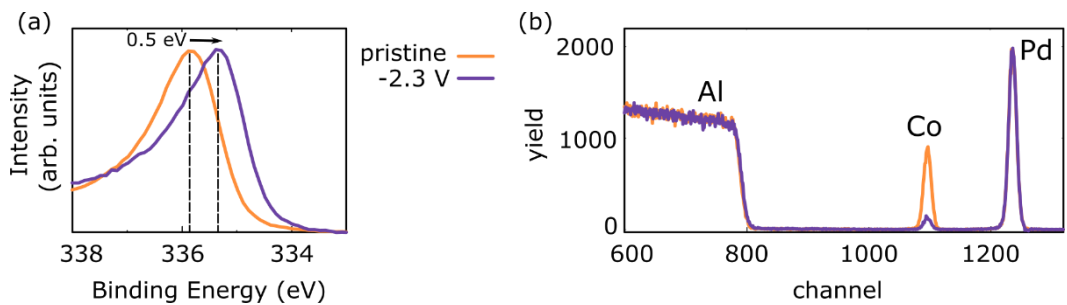

Figure 5: (a) XPS shows that Pd $3d_{5/2}$ spectrum shifts by about 0.5 eV to lower binding energies with an applied voltage of -2.3V, showing the reduced nature of the film treated at -2.3 V for 60 min. (b) RBS for the same film, showing significant loss of Co content after the electrolysis.

## 4. Conclusions

In conclusion, this work provides a way to change the nonmagnetic $PdCoO_2$ film into a reduced ferromagnetic material exhibiting perpendicular magnetic anisotropy. Above a threshold voltage of -1.2 V, the well-known water splitting potential, hydrogen is generated on the cathode and the film subsequently becomes ferromagnetic, and the anomalous Hall effect can be tuned by the voltage applied during electrolysis. This study shows that electrolysis could provide another route to manipulating magneto-electronic properties of PdCoO$_2$ films for spintronic applications.

## Acknowledgement

This work is supported by National Science Foundation, Grant No. DMR2004125 and Army Research Office (ARO), Grant No. W911NF-20-1-0108. Y.L. acknowledges the state scholarship provided by the China Scholarship Council.

List of Figures

**Figure 1:** Experiment setup. (a) Schematic diagram of the setup; (b) Underwater part of the film. Also shown is the formation of hydrogen bubbles on the film surface when a voltage of -5.0 V is applied.

**Figure 2:** Average two-point resistance ($R_{2prb}$) measured after a sample is kept in the electrolyte (a) at the specified voltage for 5 minutes and (b) for the specified time at -1.0V, -1.8 V, -2.3 V and -3.5 V. Four separate samples are used for (b). The negative voltages imply that the $PdCoO_2$ film is used as a cathode.

**Figure 3:** Hall effect data ($R_{Hall} \equiv R_{xy}$) of $PdCoO_2$ thin films measured at room temperature after electrolytic processing at (a) -1.0 V, (b) -1.8 V, (c) -2.3 V and (d) -3.5 V.

**Figure 4:** Hall resistance of $PdCoO_2$ films electrolytically processed for different duration at (a) -1.8 V, (b) -2.3 V, and (c) -3.5 V, measured at room temperature and 7 K.

**Figure 5:** (a) XPS shows that Pd $3d_{5/2}$ spectrum shifts by about 0.5 eV to lower binding energies with an applied voltage of -2.3V, showing the reduced nature of the film treated at -2.3 V for 60 min. (b) RBS for the same film, showing significant loss of Co content after the electrolysis.